\documentclass[final,3p,twocolumn,authoryear]{elsarticle}
\usepackage[hidelinks]{hyperref}
\usepackage{amssymb}
\usepackage{amsmath}
\usepackage{booktabs}

\renewcommand{\d}[2][]{\operatorname{d}^{#1}\!{#2}}

\journal{Journal of Computational Physics}

\begin{document}

\begin{frontmatter}

\title{The Runge-Kutta-Wentzel-Kramers-Brillouin method}

\author[ap,kavli]{W.J.\ Handley}%
\ead{wh260@cam.ac.uk}
\author[ap,kavli]{A.N.\ Lasenby}%
\ead{a.n.lasenby@mrao.cam.ac.uk}
\author[ap]{M.P.\ Hobson}%
\ead{mph@mrao.cam.ac.uk}

\address[ap]{Astrophysics Group, Cavendish Laboratory, J.J.Thomson Avenue, Cambridge, CB3 0HE, UK}
\address[kavli]{Kavli Institute for Cosmology, Madingley Road, Cambridge, CB3 0HA, UK}

\begin{abstract}
  We demonstrate the effectiveness of a novel scheme for numerically solving linear differential equations whose solutions exhibit extreme oscillation. We take a standard Runge-Kutta approach, but replace the Taylor expansion formula with a Wentzel-Kramers-Brillouin method. The method is demonstrated by application to the Airy equation, along with a more complicated burst-oscillation case. Finally, we compare our scheme to existing approaches.
\end{abstract}


\end{frontmatter}

\section{Introduction}
\label{sec:introduction}
The numerical solution of linear, ordinary differential equations is of critical importance throughout science and mathematics. In this paper we suggest an efficient approach for numerically solving equations with highly oscillatory solutions.

Most traditional numerical solvers of differential equations use a generalisation of Runge-Kutta (RK) techniques~\citep{ButcherMono}. These apply Taylor's theorem to create a stepping scheme whereby the value of the solution is updated using derivative information. Good solvers will also incorporate adaptive step-size control.
Whilst RK techniques are an excellent workhorse for solving a wide variety of problems, they are known to struggle to solve equations with highly oscillatory solutions. 

On the other hand, the Wentzel-Kramers-Brillouin (WKB) method is a well established analytical approach for approximately describing oscillatory solutions~\citep{RHB,Bender+2010}. Historically WKB has been used to approximate the global shape and characteristics of an oscillatory solution with a ``slowly changing'' frequency.

We propose that one may combine the two approaches to create a reliable general tool for the numerical solution of oscillatory differential equations, and term the result RKWKB\footnote{Readers with experience in the field will note that, as Cambridge authors, we should be insisting on an additional `J' in WKB (Jeffreys). Given the length of our proposed initialism, we have opted to use the more efficient nomenclature.}.

We note that the approach advocated in this paper is similar to the work of~\cite{Iserles02globalerror,Iserles01thinkglobally}, and explore the similarities and differences in Section~\ref{sec:iserles_comparison}.

\section{Background}
\subsection{Oscillatory solutions}
\begin{figure}
  \centering
  \includegraphics[width=\columnwidth]{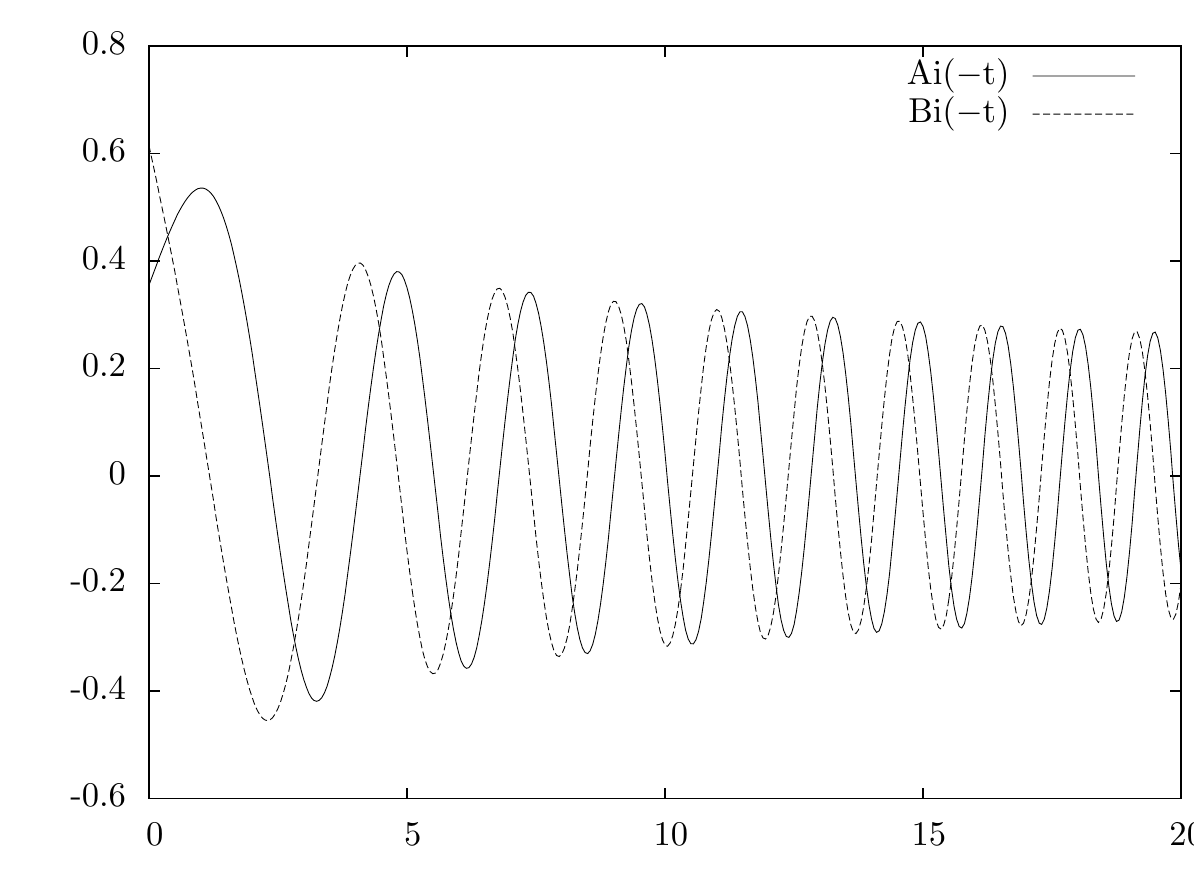}
  \caption{The real and imaginary parts of the function \(\mathrm{Ai}(-t) + \mathrm{Bi(-t)} i\), where \(\mathrm{Ai}\) and \(\mathrm{Bi}\) are the Airy functions of the first and second kind.}\label{fig:airy}
\end{figure}

We seek to create a numerical method which efficiently solves the {\em linear oscillator equation\/}:
\begin{equation}
  \ddot{x}(t) + {\omega(t)}^2x(t) = 0,\qquad \omega(t)\in\mathbb{R}.
  \label{eqn:lode}
\end{equation}
If \(\omega(t)=\omega = \mathrm{constant}\), then the solutions are sinusoidal: \(x\propto \exp{(\pm i \omega t)}\). If \(\omega(t)\) changes slowly with \(t\), then the solutions are approximately sinusoidal with a slowly varying frequency and amplitude (this statement will be made more concrete in Section~\ref{sec:wkb}). An example of such a solution can be seen in Figure~\ref{fig:airy}.

In general, any second order linear differential equation may be transformed into the form of equation~\eqref{eqn:lode} by either changing the independent variable \(t\) or dependent variable \(x\). The method we will describe can easily be adapted to other linear differential equations, but we will work with equation~\eqref{eqn:lode} for its simplicity of exposition.

Equation~\eqref{eqn:lode} is ubiquitous in physics, particularly in quantum mechanics. The authors' particular interest in its efficient solution comes from work in quantum fields in curved spacetime.

Over the next two subsections we will review the traditional techniques available for solving equations such as the linear oscillator~\eqref{eqn:lode}.

\subsection{Runge-Kutta theory}
\label{sec:rk}
We briefly review the theory of numerically solving ordinary differential equations, before discussing why Runge Kutta techniques are an inefficient tool for solving equations such as the linear oscillator~\eqref{eqn:lode}.
For a more detailed introduction to the theory of the numerical solution of ordinary differential equations we recommend~\cite{ButcherNumerical} or~\cite{hairer2008solving}, whilst a more practically oriented guide is provided by~\cite{Press+2007}.

A general non-linear differential equation in \(n\) variables can be written in terms of vectors as:
\begin{equation}
  \dot{\mathbf{y}}(t) = \mathbf{f}(\mathbf{y}(t),t).
  \label{eqn:ode}
\end{equation}
Note that any higher order differential equation can be re-written in the above form by introducing new variables for each of the higher derivative terms.

Runge-Kutta methods work effectively by generalising the Taylor expansion:
\begin{equation}
  \mathbf{y}(t+h)  = \mathbf{y}(t) + h\:\mathbf{f}(\mathbf{y}(t),t) + \mathcal{O}(h^2).
  \label{eqn:euler}
\end{equation}
Given the value of a solution \(\mathbf{y}_j\) at some time \(t_j\), one may advance to the value of the solution \(\mathbf{y}_{j+1}\) at some finite time later \(t_{j+1} = t_j + h\) by using the recursion relation:
\begin{align}
  \mathbf{y}_{j+1} &=  \mathbf{y}_{j} + h\:\mathbf{f}(\mathbf{y}_j,t_j),
  \label{eqn:RK_y_step}\\
  t_{j+1} &=  t_{j} + h.
  \label{eqn:RK_t_step}
\end{align}
This is termed {\em Euler's method}, and for arbitrarily small \(h\) will recover the solution to any desired accuracy. It is termed {\em first order\/} since each step is accurate to \(\sim\mathcal{O}(h)\).

Euler's method is normally impractical for real numerical work. Runge-Kutta schemes work by generalising~\eqref{eqn:euler}~\&~\eqref{eqn:RK_y_step} by including additional intermediate function evaluations that integrate~\eqref{eqn:ode} with greater accuracy.

A possibly more important adjustment is to equip the algorithm with the ability to choose the step size \(h\) according to the accuracy required. A popular stratagem is to run two steps, one of order \(p\), and another of order \(p-1\), and use the difference between the two as an estimate of the error. Particularly smart algorithms use the same function evaluations for both orders, an example of which is the Runge-Kutta-Fehlberg \(4(5)\) method detailed in Appendix~\ref{sec:rkf}.

All methods based on RK principles struggle to solve equations such as the linear oscillator~\eqref{eqn:lode} when the algorithm must scale a very large number of peaks and troughs. Errors accumulate rapidly in these approaches, even if the variation of \(\omega(t)\) in \(t\) is very simple. Given the regularity of the solution from Figure~\ref{fig:airy}, one would imagine that there should be a more efficient method.

\subsection{WKB theory}
\label{sec:wkb}
WKB approaches are designed to solve linear the linear oscillator equation~\eqref{eqn:lode} in the limit of a slowly varying \(\omega(t)\): i.e.\ the fractional change in frequency \(\frac{\Delta\omega}{\omega}\) over several time periods \(\Delta t \sim \frac{2\pi}{\omega}\) is relatively small.
A systematic way of phrasing ``slowly varying'' is to rescale the independent variable so \(t\rightarrow t/T\):
\begin{equation}
  \ddot{x}(t) + T^{-2}{\omega(t)}^2x(t) = 0,\qquad \omega(t)\in\mathbb{R}.
  \label{eqn:lode_T}
\end{equation}
If \(T\gg1\) then \(\omega\) is slowly varying as \(\omega\) must be large in comparison to the timescale on which \(x\) is accelerating. One can then expand the solutions in terms of complex exponential functions:
\begin{equation}
  x(t)\sim \exp\left( \frac{1}{T}\sum\limits_{n=0}^{\infty} S_n(t)\: T^n \right).
  \label{eqn:asymp}
\end{equation}
Substituting the above into equation~\eqref{eqn:lode_T} and setting each coefficient of \(T\) equal to zero yields a sequence of solvable equations. One finds the first four solutions are:
\begin{align}
  S_0(t) &= \pm i \int^t \omega(\tau)\: \d{\tau},
  \label{eqn:S0}\\
  S_1(t) &= -\frac{1}{2}\log \omega(t),\\
  S_2(t) &=  \mp i \int^t \frac{1}{4}\frac{\ddot{\omega}(\tau)}{\omega^{2}(\tau)} - \frac{3}{8}\frac{\dot{\omega}^2(\tau)}{\omega^{3}(\tau)}\: \d{\tau}, \\
  S_3(t) &=  \frac{1}{8}\frac{\ddot{\omega}(t)}{\omega^{3}(t)} - \frac{3}{16} \frac{\dot{\omega}^{2}(t)}{\omega^{4}(t)},
\end{align}
and in general:
\begin{equation}
    \dot{S}_0(t) = \pm i \omega, \qquad \dot{S}_n = -\frac{1}{\dot{S}_0} \left( \ddot{S}_{n-1}+ \sum_{j=1}^{n-1}\dot{S}_j\dot{S}_{n-j}  \right).
    \label{eqn:WKB_general}
\end{equation}
Note that at \(0\)\textsuperscript{th} order, the solution has \({x\propto\exp\left(\pm i \int \omega \d{t} \right)}\), which should be compared with the traditional sinusoidal solution.
Typically \(T\) is considered a power counting parameter, and set equal to \(1\) at the end of the analysis.
For further detail on the intricacies of WKB approaches, the reader should consult~\cite{RHB,Bender+2010}.

\section{Generalised stepping methods}
\label{sec:stepping_methods}
In Section~\ref{sec:rkwkb} we will combine the power of WKB approaches with RK to form a method specialised in navigating oscillatory solutions. However, to avoid confusing the intricacies of WKB with the generic nature of our stepping methodology, we will work with a general stepping function for this section. For concreteness we will focus on second order differential equations, and discuss the extension to higher orders at the end of this section.

We aim to solve a second order differential equation in \(x(t)\) for which we have some class of functions that form an approximate basis set for the true solutions.\footnote{For example, the linear oscillator equation~\protect\eqref{eqn:lode}, with approximate WKB solutions~\protect\eqref{eqn:asymp}--\protect\eqref{eqn:WKB_general}.} 
For the case of second order differential equations, one requires two approximate solutions \(f_\pm\) that remain independent for all \(t\).

\begin{figure}
  \includegraphics[width=\columnwidth]{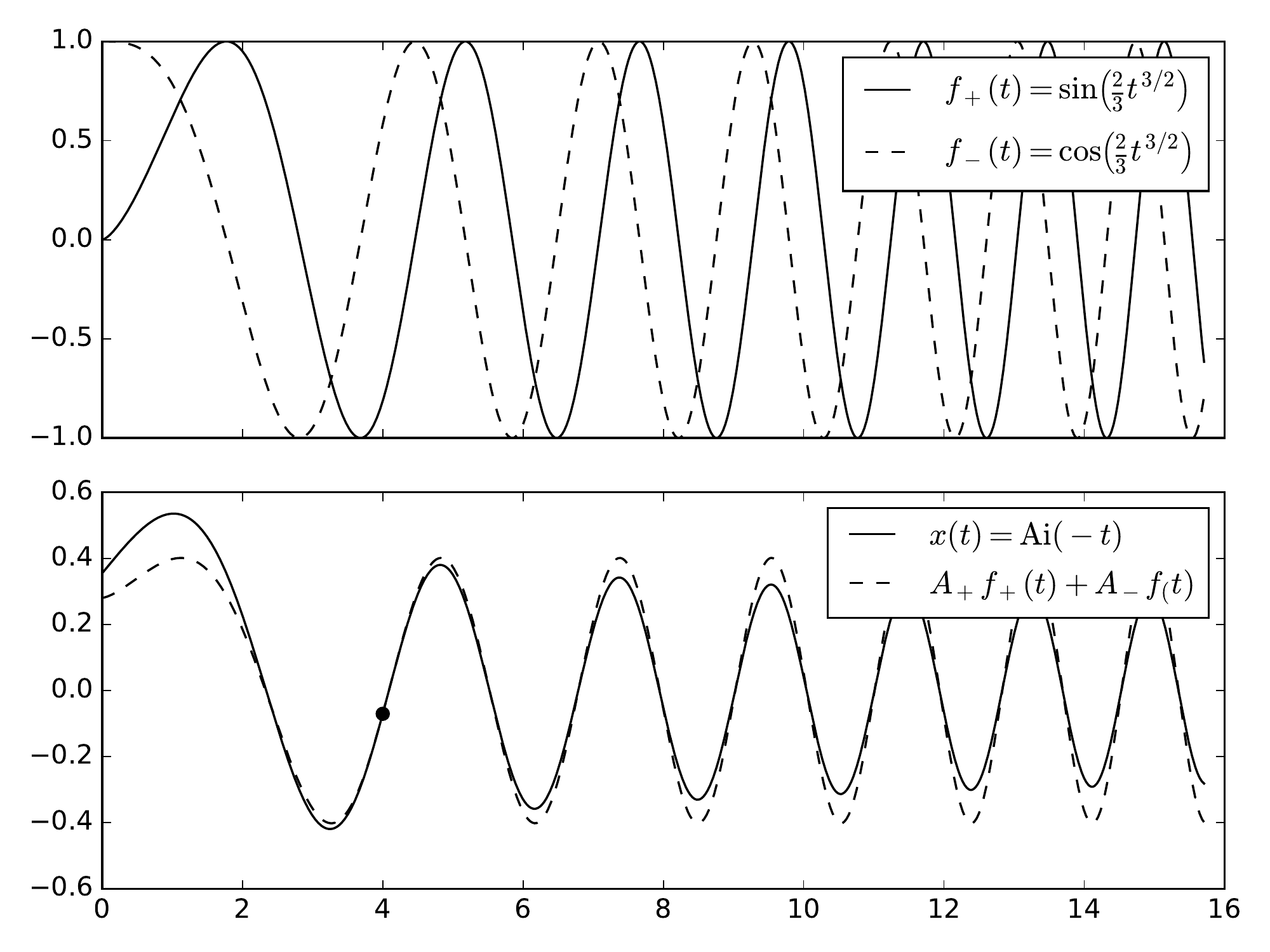}
  \caption{Approximating an Airy function with two sinusoids of varying frequency. The lower figure shows the Airy function of the first kind \(\mathrm{Ai}(-t)\) (solid line) being approximated in the region around \(t_j=4\) (dashed line). Sinusoids of varying frequency are matched onto the Airy function's value and derivative at \(t_j\). The top part of the figure details these sinusoids, which are equivalent to 0\textsuperscript{th} order WKB solutions.}\label{fig:matching}
\end{figure}

With these two solutions in hand, at any given time \(t_j\) with values of the true solution \(x_j\) and its derivative \(\dot{x}_j\), one may match the approximate solutions onto the correct solution:
\begin{align}
    x(t) &\approx  A_+ f_+(t) + A_- f_-(t) ,
    \label{eqn:x_approxf}\\
    A_\pm &= \frac{\dot{x}_j f_\mp(t_j) - x_j \dot{f}_\mp(t_j) }{\dot{f}_\pm(t_j) f_\mp(t_j) - \dot{f}_\mp(t_j) f_\pm(t_j)}.
    \label{eqn:Apm}
\end{align}
Equation~\eqref{eqn:x_approxf} provides an approximation to the true solution in the region \(t=t_j + h\) where \(h\) is small, an example of which may be seen graphically in Figure~\ref{fig:matching}.

A na\"{\i}ve approach would be to use the approximate matched solution~\eqref{eqn:x_approxf} to create a stepping procedure analogous to the RK method~\eqref{eqn:RK_y_step}~\&~\eqref{eqn:RK_t_step}:
\begin{align}
  x_{j+1} &= A_+ f_+(t_j + h) + A_- f_-(t_j + h) ,\label{eqn:x_step_wrong}\\
  \dot{x}_{j+1} &= A_+ \dot{f}_+(t_j + h) + A_- \dot{f}_-(t_j + h) ,\label{eqn:x_dot_step_wrong}\\
  t_{j+1} &= t_j+h\label{eqn:t_step_wrong}.
\end{align}
Alas, such a method is doomed to failure\footnote{Credit here is due to Anthony Challinor for spotting this error in an earlier version of our approach.}. Since the approximate solution \(A_+ f_+ + A_- f_-\) is defined entirely by the value of \(x\) and \(\dot{x}\) at any given point, using the values \(x_j\) and \(\dot{x}_{j}\) to forecast onto \(x_{j+1}\) and \(\dot{x}_{j+1}\) merely continues the solution of the previous step. The coefficients \(A_\pm\) do not change, and such a method merely follows a single curve ad infinitum.

One can see the failure more concretely by observing that any convergent method should replicate a simple Runge-Kutta approach in the limit of vanishing step size \(h\). In the limit of small \(h\), one has:
\begin{align}
    f(t_{j+1}) &\approx f(t_j) + \dot{f}(t_j) \: h +\mathcal{O}(h^2), \\
    \dot{f}(t_{j+1}) &\approx f(t_j) + \ddot{f}(t_j) \: h +\mathcal{O}(h^2). 
\end{align}
Substituting these into equations~\eqref{eqn:x_step_wrong}~\&~\eqref{eqn:x_dot_step_wrong} yields:
\begin{align}
    \Rightarrow x_{j+1} &= x_j + \dot{x}_j \: h , \\
    \Rightarrow \dot{x}_{j+1} &= \dot{x}_j + 
    \frac{(\ddot{f}_+ f_- - \ddot{f}_- f_+)\dot{x}_j  +( \ddot{f}_- \dot{f}_+- \ddot{f}_+ \dot{f}_-)x_j}{\dot{f}_+ f_- - \dot{f}_- f_+}\: h
    , 
\end{align}
where in the final equation, all \(f\) terms are evaluated at \(t_j\). In general, unless \(f\) is an exact solution, one cannot expect the coefficient of \(h\) in the second equation to be the same as \(\ddot{x}_j\), and thus the approach defined by equations~\eqref{eqn:x_step_wrong}--\eqref{eqn:t_step_wrong} fails to recover the Runge-Kutta result. Thus, decreasing the step size does not improve accuracy, and the algorithm is not convergent.

The above issue also suggests a solution. We require an alternative to~\eqref{eqn:x_dot_step_wrong} which steps \(\dot{x}_j\) such that in the limit of small \(h\) the stepping procedure reduces to \({\dot{x}_{j+1} = \dot{x}_j + \ddot{x}_j\: h}\).
We should therefore perform a separate step for \(\dot{x}\), with the solution matched onto the values of \(\dot{x}_j\) and \(\ddot{x}_j\):
\begin{align}
    \dot{x}(t) &\approx  B_+ \dot{f}_+(t) + B_- \dot{f}_-(t), \\
    B_\pm &= \frac{\ddot{x}_j \dot{f}_\mp(t_j) - \dot{x}_j \ddot{f}_\mp(t_j) }{\ddot{f}_\pm(t_j) \dot{f}_\mp(t_j) - \ddot{f}_\mp(t_j) \ddot{f}_\pm(t_j)}. 
    \label{eqn:Bpm}
\end{align}
Most importantly, one may determine \(\ddot{x}_j\) from \(\dot{x}_j\) and \(x_j\) via the original second order differential equation.

The general stepping procedure is then as follows:
\begin{align}
  x_{j+1} &= A_+ f_+(t_j + h) + A_- f_-(t_j + h),
  \label{eqn:x_step} \\
  \dot{x}_{j+1} &= B_+ \dot{f}_+(t_j + h) + B_- \dot{f}_-(t_j + h),
  \label{eqn:x_dot_step} \\
  t_{j+1} &= t_j+h,
  \label{eqn:t_step}
\end{align}
and at every iteration, \(A_\pm\) and \(B_\pm\) are determined by equations~\eqref{eqn:Apm}~\&~\eqref{eqn:Bpm} with \(\ddot{x}_j\) calculated from \(\dot{x}_j\) and \(x_j\) via the original second order linear differential equation. By definition (or somewhat laborious algebra), in the limit of small \(h\), equations~\eqref{eqn:x_step}--\eqref{eqn:t_step} recover the RK result:
\begin{align}
    x_{j+1} &= x_j + \dot{x}_j \: h + \mathcal{O}(h^2), \\
    \dot{x}_{j+1} &= \dot{x}_j  + \ddot{x}_j\: h + \mathcal{O}(h^2). 
\end{align}

The procedure outlined in this section may easily be extended to \(n\)\textsuperscript{th} order differential equations. One simply requires \(n\) independent approximate solutions \(\{f_1,\ldots,f_n\}\), along with \(n\) analogues of the stepping equations~\eqref{eqn:x_step}~\&~\eqref{eqn:x_dot_step} for each of the derivatives \(\{x,\dot{x},\ldots,x^{(n-1)}\}\) using an \(n\times n\) matrix of coefficients analogous to \(\left(\begin{array}{ll}A_+ &A_-\\ B_+ & B_-\end{array}\right)\).

Furthermore, whilst we will spend the remainder of this paper focussing on using WKB approximate solutions for the linear oscillator equation~\eqref{eqn:lode}, the generalised stepping approach would be equally valid for non-linear equations.

Finally, it is worthy of note that if one chooses \(f_\pm\) to be linear polynomials, then one recovers Euler's method~\eqref{eqn:RK_y_step}~\&~\eqref{eqn:RK_t_step}. Care must be taken with the indeterminacy of the \(B_\pm\) coefficients, but a clean derivation can be found by choosing \(f_+ = t\) and \(f_- = 1 + \epsilon t^2\), and taking \(\epsilon\to 0\) at the end of the analysis. One can therefore consider the procedure described in this section as a true generalisation of a Runge-Kutta approach, the effectiveness of which is entirely determined by how well \(f_\pm\) approximates the true solution.

\section{The RKWKB method}
\label{sec:rkwkb}
We now specialise the generic technique detailed in Section~\ref{sec:stepping_methods} to the case of equations with oscillatory solutions. Our strategy is to combine the versatility of RK methods with the power of WKB in dealing with oscillatory solutions, and term the combination RKWKB\@. 

We apply the generalised stepping technique~\eqref{eqn:x_step},~\eqref{eqn:x_dot_step}~\&~\eqref{eqn:t_step} by choosing \(f_\pm\) to be the WKB solutions:
\begin{equation}
    f_\pm(t) = \frac{1}{\sqrt{\omega(t)}}\exp\left(\pm i \int^t \omega(\tau)\d{\tau} +\ldots\right).
\end{equation}
As can be seen in Figure~\ref{fig:matching}, choosing these as \(f_\pm\) means that our stepping procedure naturally encodes the oscillatory nature of the solutions, particularly if the frequency is large. Instead of following every peak and trough as a RK scheme must do, it is potentially able to leap over many oscillations at once, greatly increasing the speed and accuracy of the numerical solution.

\subsection{Step size adjustment}
To tune the step size \(h\), we use the same strategy as adaptive Runge-Kutta schemes. We compute both the order \(n\) and order \(n-1\) WKB solutions, and use the fractional difference between the two:
\begin{equation}
  \varepsilon = \left|\frac{x^{(n)}-x^{(n-1)}}{x^{(n)}}\right|,
\end{equation}
as an estimate of the truncation error. 

We now assume that the desired accuracy is \(\alpha\). If \(\varepsilon<\alpha\) then the solution is within the desired tolerance, and the algorithm makes a step of size \(h\). \(h\) is then increased for the next iteration. If \(\varepsilon>\alpha\) then the step is unsuccessful, and the step size is reduced. \(h\) may therefore be efficiently updated between attempts via:
\begin{equation}
  h \to h\times\left\{
  \begin{array}{lr}
    {(\alpha/\varepsilon)}^{1/n} &: \varepsilon<\alpha \\
    {(\alpha/\varepsilon)}^{1/(n-1)} &: \varepsilon>\alpha. \\
  \end{array}
  \right.\label{eqn:h_update}
\end{equation}
This updating procedure allows the step size to increase in the regions where the initial step size is unnecessarily small, whilst ensuring that the step size is always small enough to keep forecasts within a given error margin.

\subsection{Dynamic switching}
In general, one cannot expect the WKB expansion to be an accurate approximation throughout the solution region. If \(\omega\) is too small, or too quickly varying, then the step size \(h\) will decrease to an inefficiently small size. This problem can be countered by simultaneously attempting a step using a standard adaptive RK method. One chooses between RK and WKB by selecting the method with the smallest error, providing a natural switching mechanism, without having to delve into the details of the validity of the WKB expansion.

We choose the Runge-Kutta-Fehlberg \(4(5)\) method for our alternative solver, which is detailed in Appendix~\ref{sec:rkf}. However, this specific RK method may be substituted with any ODE solver according to the user's preference.

\section{Example: The Airy equation}

As an example of the RKWKB approach, we apply it to the Airy equation:
\begin{align}
  0=\ddot{x}(t) &+ t\: x(t) ,
  \label{eqn:airy_equation}\\
  x(0)=\frac{3^{-2/3}+3^{-1/6}i}{\Gamma(2/3)},
  &\qquad
  \dot{x}(0) = \frac{3^{-1/3}-3^{1/6}i}{\Gamma(1/3)},
  \\
  \Rightarrow x(t) = \mathrm{Ai}(-t) &+ \mathrm{Bi}(-t)\:i,
  \label{eqn:airy_solution}
\end{align}
whose solution is depicted in Figure~\ref{fig:airy}. The Airy equation is often quoted as being a ``maximally hard'' problem for RK machinery to solve, since the frequency steadily increases, causing the step size to get smaller as the algorithm goes deeper into the solution.

\begin{figure}
  \centering
  \includegraphics[width=\columnwidth]{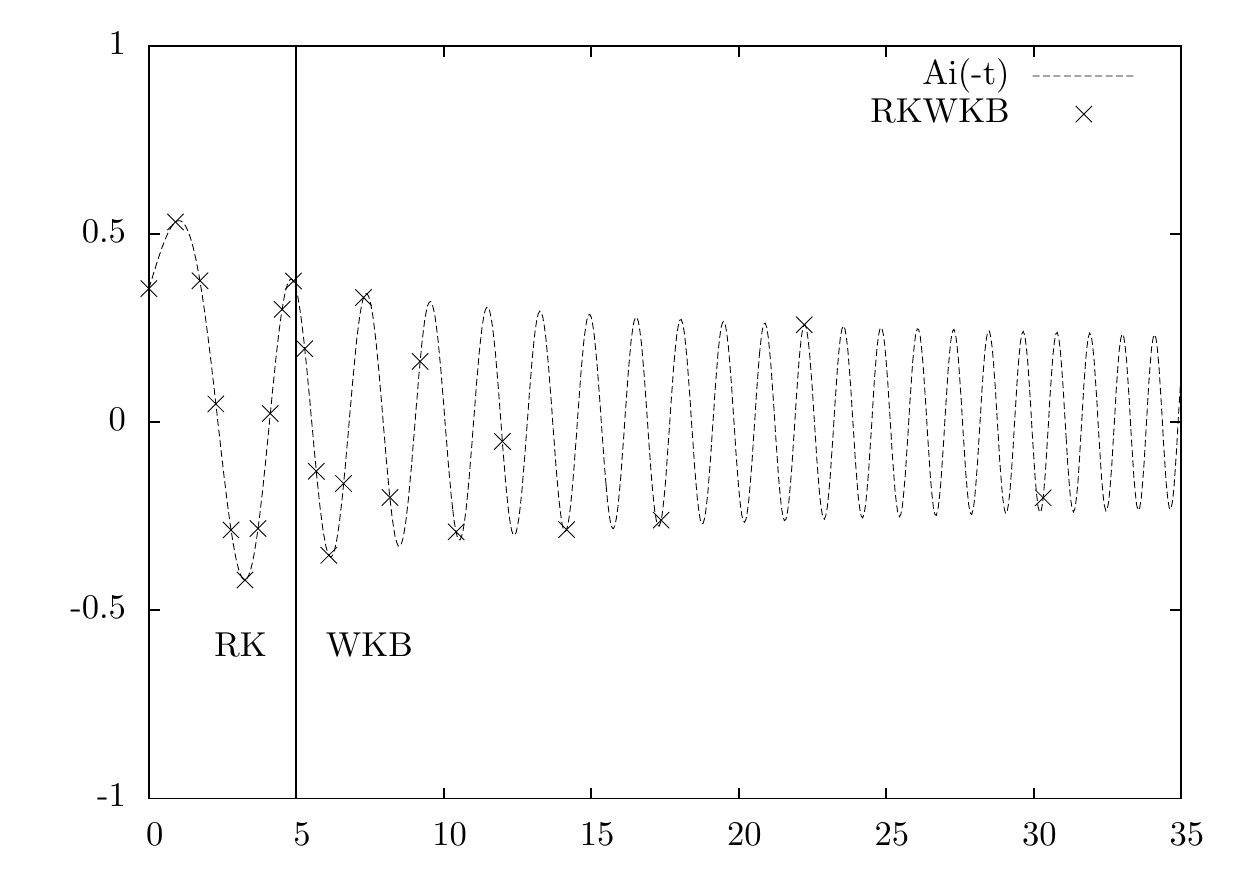}
  \caption{The RKWKB method compared with the analytical solution. The algorithm starts at \(t=0\) in the RK regime, since \(\omega\) is varying quickly relative to the oscillation period. At \(t=5\) it becomes more efficient to use the WKB regime, and the points start to increase in separation. By \(t=15\) the algorithm is skipping multiple periods, and the step size \(h\) increases exponentially.}\label{fig:data}
\end{figure}

We set the desired relative error to be \(10^{-4}\). The algorithm remains in the RK regime until \(t\sim5\). When the WKB solver is activated, instead of following every oscillation of the solution, it rapidly speeds up, skipping many oscillations. This behaviour is detailed in Figure~\ref{fig:data}.

\begin{figure}
  \centering
  \includegraphics[width=\columnwidth]{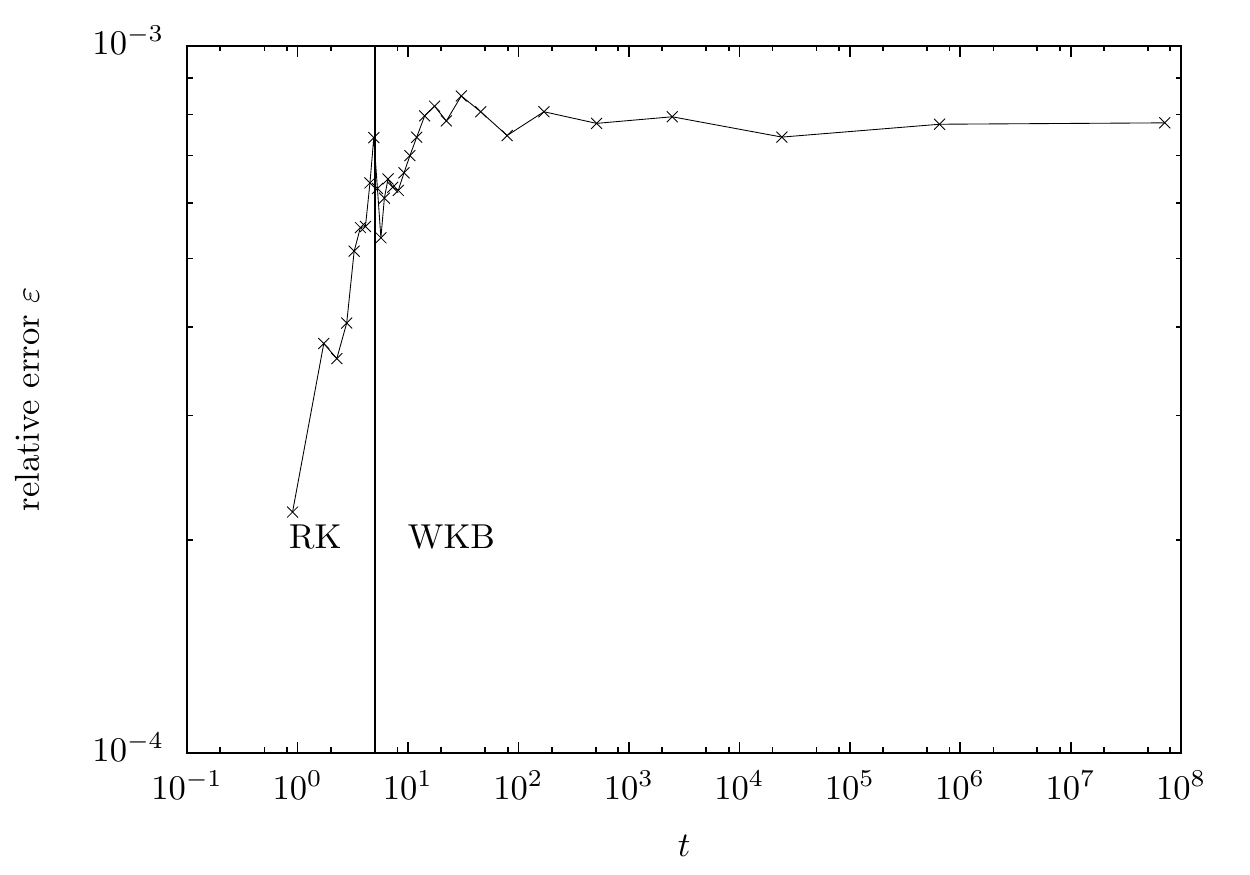}
  \caption{Fractional difference between the analytical solution and RKWKB solution from Figure~\protect\ref{fig:data}. The algorithm's fractional error begins at \(t=0\) with an error of \(\sim10^{-4}\), but rises in the RK phase. This is to be expected as RK methods accumulate errors (particularly for oscillatory solutions). Upon entering the WKB region, the fractional error levels off. Note the rapidly increasing step size, and accuracy at extremely late times \(t\). To the authors' knowledge, no numerical scheme to date has demonstrated the ability to solve the Airy equation~\protect\eqref{eqn:airy_equation} to times as late as this.}\label{fig:error}
\end{figure}

The error compared to the true solution is detailed in Figure~\ref{fig:error}. Here we find that initially the error is small, but grows\(\sim \mathcal{O}(h^{2s} t^{s+5/4})\) where \(s=4\) is the order of the RK method~\citep{Iserles02globalerror}. After the WKB regime is entered, it begins to make huge strides, and the error levels off.

In contrast to a ``pure'' RK method, the RKWKB method finds the Airy equation maximally {\em easy}.

\section{Example: Burst oscillation}
\begin{figure}
  \includegraphics[width=\columnwidth]{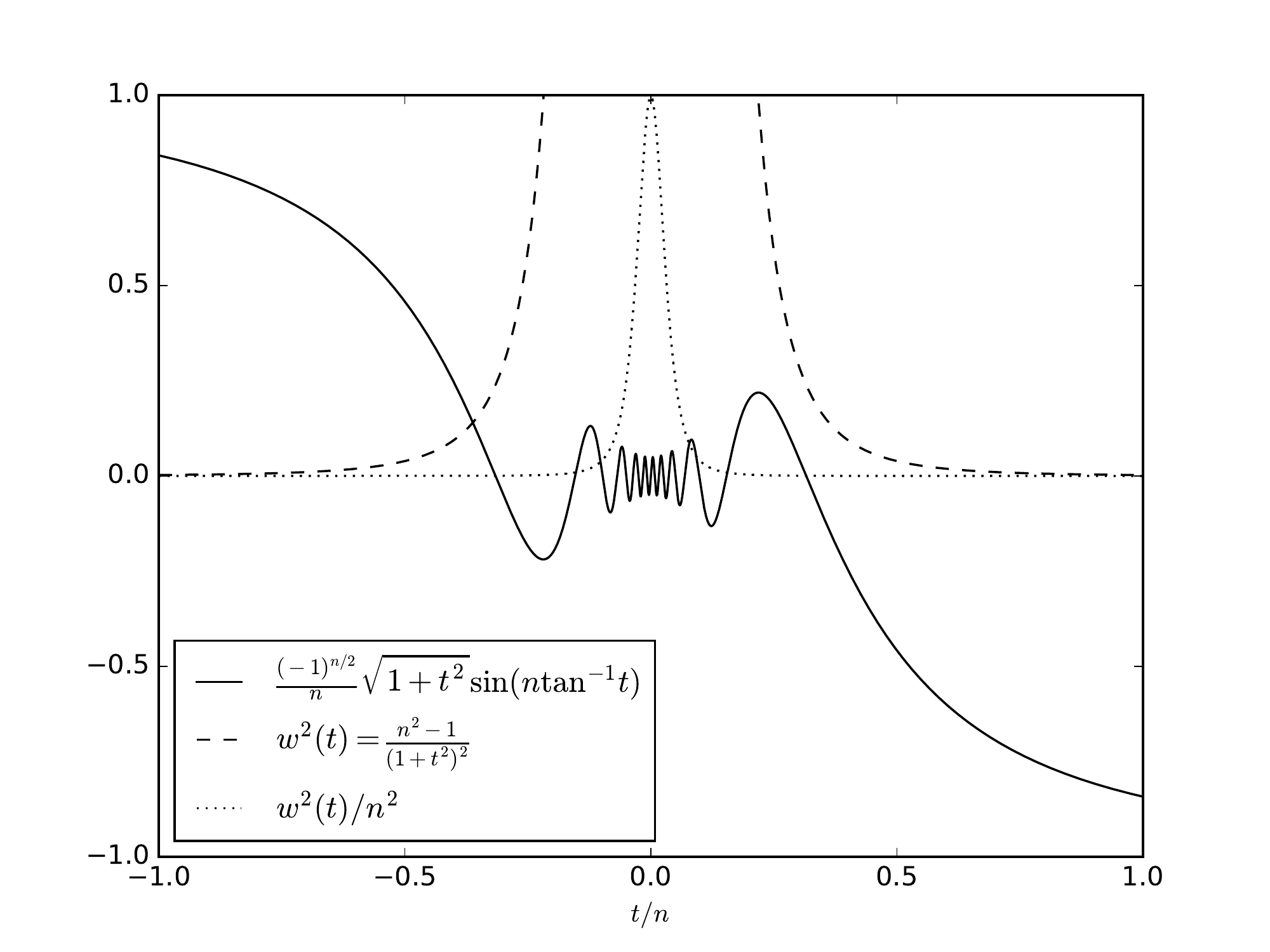}
  \caption{An example of a solution to the burst equation~\protect\eqref{eqn:burst}, with \(n=20\). Also plotted is the region in which the frequency is effectively non-zero (dashed line), and a dotted line indicating the profile of the time-dependent frequency.}\label{fig:burst}
\end{figure}

We now turn to a more complicated example. Here we aim to solve the linear oscillator equation~\eqref{eqn:lode} when:
\begin{equation}
    \omega^2(t) = \frac{n^2-1}{{(1+t^2)}^2}.
\end{equation}
As is shown in Figure~\ref{fig:burst}, this is a frequency profile which is zero everywhere apart from a region around the origin \(\sim \pm n\). Solutions will be a straight line far from the origin, but near to the origin will exhibit a burst of strong oscillations. If we choose stationary initial conditions:
\begin{align}
    \ddot{x}(t) + \frac{n^2-1}{{(1+t^2)}^2} x(t) = 0,\nonumber\\ 
    \quad x(-\infty) = 1,\quad \dot{x}(-\infty) = 0, \label{eqn:burst}
\end{align}
then for integer \(n\) the above equation solves to give:
\begin{equation}
    x(t) = \frac{\sqrt{1+t^2}}{n}\left\{
        \begin{array}{rl}
            {(-1)}^{n/2} \sin\left( n \tan^{-1}t \right) & \text{: \(n\) even}\\
            {(-1)}^{(n-1)/2} \cos\left( n \tan^{-1}t \right) & \text{: \(n\) odd,}
        \end{array}
    \right.
\end{equation}
Such solutions begin entirely flat, oscillating approximately \(n/2\) times in the region \(t\sim\pm n\), before finishing again in a flat state.

\begin{figure}
  \includegraphics[width=\columnwidth]{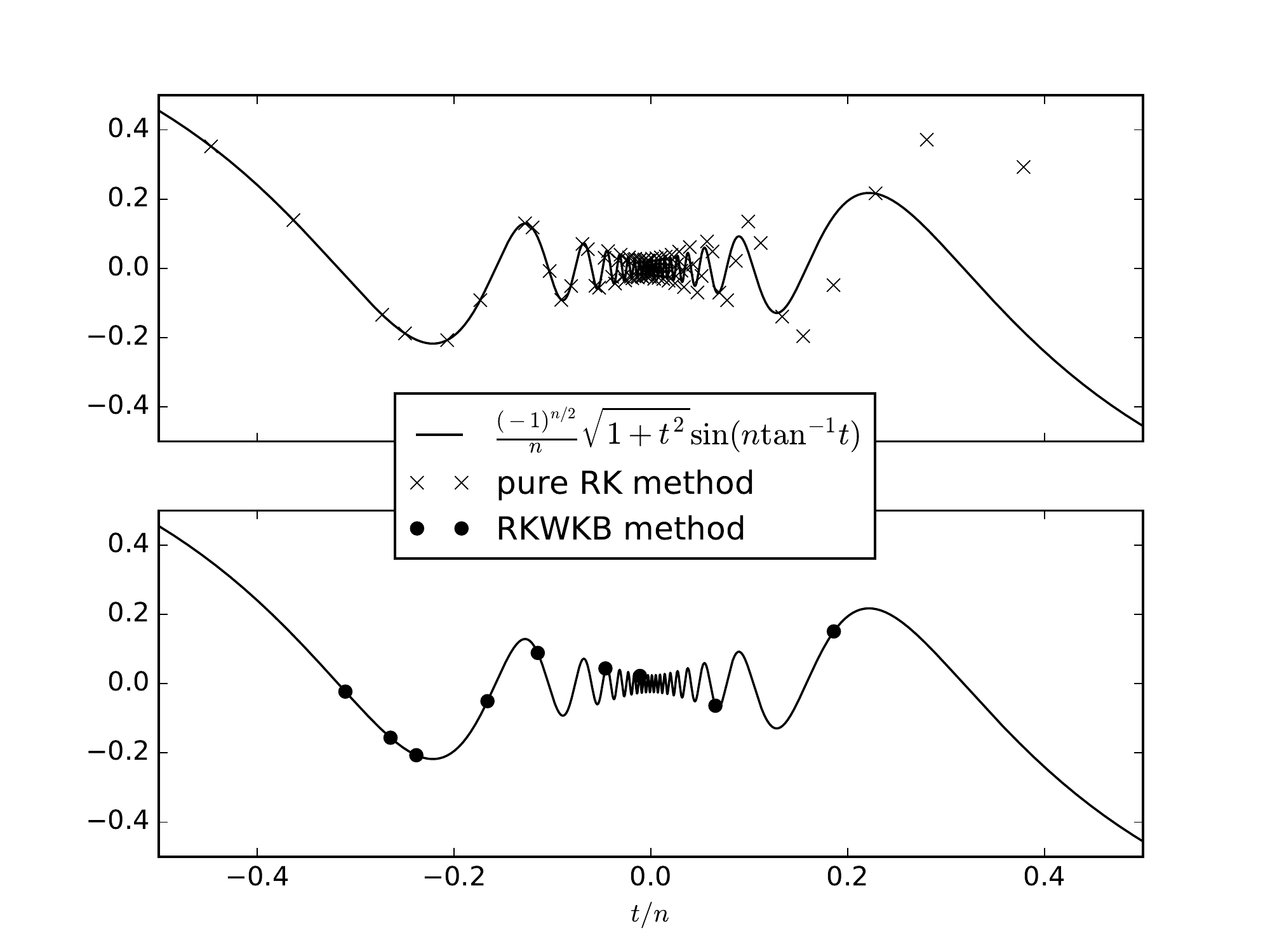}
  \caption{Numerical solutions to the burst equation~\protect\eqref{eqn:burst} with \(n=20\). The upper figure indicates the failure of a pure RK approach. After the burst of oscillation around \(t\sim\pm n\), is unable to recover the correct analytic solution. The lower figure shows that RKWKB on the other hand recovers it to within the same desired error (\(10^{-3}\)). }\label{fig:burst_compare}
\end{figure}

If one applies a simple RK scheme to the above equation, the period of rapid oscillation introduces large inaccuracies, as the stepping scheme must follow each peak and trough. This is demonstrated in the upper half of Figure~\ref{fig:burst_compare}. The RK method navigates the initial slowly-changing phase well, but encounters difficulties around the origin where the solution exhibits a burst of oscillations, causing a consequent dramatic reduction in step size. Moreover, the error accumulated around the origin means that the numerical solution no longer matches the analytic solution after leaving the oscillatory phase. The result predicted by a pure RK approach thus fails to recover the flat ending to the solution.

Applying the RKWKB approach to the same system recovers the correct solution precisely (lower half of Figure~\ref{fig:burst_compare}. On the approach to the burst, both RK and WKB have approximately the same error, so the algorithm seemingly chooses at random between the two. At the burst, it switches to a WKB stepper, and crosses the oscillations in a couple of steps. Afterwards, it switches back to the initial behaviour, but this time recovering the correct solution to within error.

It should be noted that we have chosen a relatively low \(n\) for the purposes of visualisation. If one chooses higher \(n\), then the pure RK attempt at a solution becomes far worse, whilst the RKWKB solution becomes much better, crossing the entirety of the oscillatory regime in a single stride.

\section{Comparison with the Iserles approach}
\label{sec:iserles_comparison}
Iserles has written extensively on the difficulty of solving differential equations with oscillatory solutions. His approach is to turn the linear oscillator~\eqref{eqn:lode} into a Lie-group differential equation~\citep{Iserles00lie-groupmethods} by writing:
\begin{align}
  \mathbf{y} &= {(x,\dot{x})}^\top, 
  \mathrm{A}(t) = 
  \left(
  \begin{array}{cc}
    0 & 1 \\
    -\omega^2(t) & 0
  \end{array}
  \right),
  \\
  \Rightarrow\quad 
  \dot{\mathbf{y}} &= \mathrm{A} (t) \: \mathbf{y}.\label{eqn:lie_eqn}
\end{align}
This may then be attacked with a variety of Lie group methods. For example, one may write the full solution as a Magnus expansion:
\begin{align}
  \mathbf{x}(t) &= e^{\Omega(t,t_0)} \mathbf{x}_0,
  \label{eqn:magnus}\\
  \Omega(t,t_0) &= \int_{t_0}^t \mathrm{A}(x) \d{x} \nonumber \\
  &- \frac{1}{2}\int_{t_0}^t\int_{t_0}^{x_1} \left[\mathrm{A}(x_2),\mathrm{A}(x_1)\right] \d{x_2} \d{x_1} + \ldots,
\end{align}
and then use a truncated series to create a stepping algorithm.
This approach is much improved by transferring to a fast rotating frame:
\begin{equation}
  \mathbf{y}(t_n+\tau) = e^{\tau \mathrm{A}(t_n+h/2)} \mathbf{x},
  \label{eqn:rotating_frame}
\end{equation}
the end product is then termed the {\em modified Magnus method\/}~\citep{Iserles01thinkglobally}.

The RKWKB method and the modified Magnus method share some key features. Indeed, the lowest order modified Magnus method is equivalent to a \(1\)\textsuperscript{st} order WKB approach~\citep{Iserles02globalerror}. However, our approach is distinguished in several ways. 

First, the Magnus expansion~\eqref{eqn:magnus} requires multiple integrals for higher order terms, which can be tricky to implement. The WKB expansion~\eqref{eqn:WKB_general} on the other hand requires at most single integrals, replacing the double integrals with additional derivative terms of \(\omega\), which are typically easier to work with.

Second, our approach uses adaptive step-size control, which is very easy to implement in the WKB framework and crucial for real-world numerical work.

Finally, by using dynamic switching, the algorithm is able to utilise the optimal approach in real time.

However, Iserles' approach has been the inspiration for this work, and it is possible that many of the difficulties associated with the implementation of Magnus methods are merely engineering problems. This could mean that in the fullness of time Magnus methods could become the de-facto numerical integration tool. In the mean-time, this work provides a simpler, more streamlined methodology.

\section{Conclusions}

We have presented a novel method for numerically solving linear differential equations with highly oscillatory solutions. We use a Wentzel-Kramers-Brillouin expansion to create an adaptively stepping algorithm in the same manner as a Runge-Kutta scheme. Further, the algorithm will switch back to a normal RK approach when the frequency of oscillation is varying too quickly for WKB to approximate accurately. The RKWKB method is compared to Iserles existing approaches, and found to be a reasonable alternative without requiring the use of heavy Lie-group machinery. This paper is not intended to be a complete exposition, but more a proof-of-principle to create a springboard for further investigation.

\section*{Acknowledgements}
The authors thank Arieh Iserles for a very profitable discussion which formed the inspiration for this paper, as well as Anthony Challinor for spotting an error in an earlier version of the algorithm. Will Handley was partially supported by the European Research Council under the European Community's Seventh Framework Programme (FP7/2007--2013)/ERC grant agreement no.~306478--CosmicDawn, and by STFC.

\begin{table}
  \centering
\begin{equation*}
  \begin{array}{l | c c c c c}
    0      \quad &               &              &              &         &   \\
    c_2    \quad & \quad a_{21}  &              &              &         &   \\
    c_3    \quad & \quad a_{31}  & \quad a_{32} &              &         &   \\
    \vdots \quad & \quad \vdots  & \quad \vdots & \quad \ddots &         &   \\
    c_s    \quad & \quad a_{s1}  & \quad a_{s2} & \quad \cdots & \quad a_{s,s-1} & \\ \hline
    & \quad b_{1}   & \quad b_{2}  & \quad \cdots & \quad b_{s-1}  & \quad b_{s}
  \end{array}
\end{equation*}
  \caption{Butcher tableau for a general explicit RK method.}\label{tab:RKexplicit}
\end{table}

\begin{table}
  \centering
  \begin{equation*}
    \begin{array}{l | c c c c c c}
      0      &           &            &             &             &        &\\
      \frac{1}{4}    & \frac{1}{4}       &            &             &             &        &\\
      \frac{3}{8}    & \frac{3}{32}      & \frac{9}{32}       &             &             &        &\\
      \frac{12}{13}  & \frac{1932}{2197} & -\frac{7200}{2197} & \frac{7296}{2197}   &             &        &\\
      1      & \frac{439}{216}   & -8         & \frac{3680}{513}    & -\frac{845}{4104}   &        &\\
      \frac{1}{2}    & -\frac{8}{27}     & 2          & -\frac{3544}{2565}  & \frac{1859}{4104}   & -\frac{11}{40} &\\\hline
      & \frac{25}{216}    & 0          & \frac{1408}{2565}   & \frac{2197}{4104}   & -\frac{1}{5}   & 0      \\
      & \frac{16}{135}    & 0          & \frac{6656}{12825}  & \frac{28561}{56430} & -\frac{9}{50}  & \frac{2}{55}
    \end{array}
  \end{equation*}
  \caption{Butcher tableau for the embedded Runge-Kutta-Fehlberg \(4(5)\) method.}\label{tab:rkf45}
\end{table}
\section*{References}

\bibliographystyle{elsarticle-harv} 
\bibliography{RKWKB}

\appendix
\section{Runge-Kutta-Fehlberg}
\label{sec:rkf}

A general explicit RK method can be written as:
\begin{align}
  y_{n+1} &= y_n + h\sum\limits_{i=1}^{s} b_i k_i, \label{eqn:rk_step} \\
  k_s &= f(t_n + c_s h, y_n + h \sum\limits_{i=1}^{s-1}a_{si} k_i),
\end{align}
where the coefficients \(\{c_i,a_{si}\}\) are determined by the choice of method and are typically written in a Butcher tableau (Table~\ref{tab:RKexplicit}). 

A particularly efficient example is the Runge-Kutta-Fehlberg \(4(5)\) method which uses an embedded approach. It performs a fourth order step and a fifth order step, and uses the difference between these as an estimate of the error. Impressively, both steps are calculated using the same values of \(\{k_i\}\) (but different values of \(\{b_i\}\)), and hence the method only requires five function evaluations of \(f\) per step. Its Butcher tableau is detailed in Table~\ref{tab:rkf45}.

\end{document}